# Weak low-temperature ferromagnetism and linear magnetoresistance in $Lu_{0.75}Fe_6Sn_6$ with a disordered $HfFe_6Ge_6$-type structure

Chenfei Shi[1], Zhaodi Lin[1], Qiyuan Liu[1], Junai Lv[2], Xiaofan Xu[3], Baojuan Kang[1], Jin-Hu Yang[3], Yi Liu[4], Jian Zhang[2], Shixun Cao[1, 5*], and Jin-Ke Bao[1, 3, 5*]

[1] *Department of Physics, Materials Genome Institute and International Center for Quantum and Molecular Structures, Shanghai University, Shanghai 200444, People's Republic of China*

[2] *State Key Laboratory of Crystal Materials, Shandong University, Jinan 250100, People's Republic of China*

[3] *School of Physics and Hangzhou Key Laboratory of Quantum Matters, Hangzhou Normal University, Hangzhou 311121, People's Republic of China*

[4] *Key Laboratory of Quantum Precision Measurement of Zhejiang Province, School of Physics, Zhejiang University of Technology, Hangzhou 310023, China*

[5] *Shanghai Key Laboratory of High Temperature Superconductors, Shanghai University, Shanghai 200444, People's Republic of China*

*Corresponding Authors: Shixun Cao (sxcao@shu.edu.cn), and Jin-Ke Bao (baojk7139@gmail.com)

**Abstract:** We report the synthesis of $Lu_{0.75}Fe_6Sn_6$ single crystals with a Fe-kagome lattice using a self-flux method. The crystal structure, magnetic, thermodynamic and electrical transport properties were investigated. Structure refinement reveals that $Lu_{0.75}Fe_6Sn_6$ has a $HfFe_6Ge_6$-type structure as the major framework intergrown with a CoSn-type structure, leading to a vacancy of 25% on the Lu-site and disorder on the Sn-site. It exhibits a significant magnetic anisotropy with weak ferromagnetism in the *ab*-plane below 40 K and antiferromagnetic behavior along the *c*-axis. The weak ferromagnetism is due to the canted antiferromagnetism with magnetic moment deviating from the *c*-axis to the *ab*-plane. Besides, an anisotropic non-saturated linear magnetoresistance is also observed in $Lu_{0.75}Fe_6Sn_6$, probably resulting from the structural disorder in the sample.

magnetoresistance

1. **Introduction**

Kagome materials have generated extensive attention due to the unique geometric frustration from the corner-shared triangles in kagome lattice which can support Dirac points, flat bands, and van Hove singularities (vHSs) in their electronic bands, thus giving birth to various quantum phenomena [1-3] such as topological Hall effect [4,5], Weyl Fermi arc [6], giant anomalous Hall effect [7], quantum spin liquid behavior [8-10]. Recently, kagome materials are demonstrated to be a suitable platform to realize exotic quantum states and study their interactions among the following various structure types: a time-reversal symmetry breaking charge density wave (CDW) order accompanied by superconductivity has been found in the nonmagnetic kagome family $AV_3Sb_5$ ($A$ = K, Rb and Cs) [11]; the isostructural compounds $CsCr_3Sb_5$ exhibits concurrent CDW and antiferromagnetic (AFM) orders which can be suppressed to achieve superconductivity under high pressure [12] while $CsTi_3Bi_5$ presents electronic nematicity without CDW [13]; the existence of CDW under the background of the A-type AFM order in the CoSn-type FeGe triggers the enhanced magnetic moment of Fe and the dimerization of Ge [14-16], indicating the intricate interactions between charge, spin and lattice; CDW in $ScV_6Sn_6$ manifests diffuse X-ray scattering in a **q**-vector of (1/3,1/3,1/2) above $T_{CDW}$ but leads to structural distortion with a **q**-vector of (1/3,1/3,1/3) below $T_{CDW}$ [17,18].

Among the kagome materials, $AM_6X_6$ ($A$ = Mg, Ti, Zr, Hf and rare earths; $M$ = V, Cr, Mn, Fe, Co, $X$ = Ge, Sn) ("166" family) has attracted substantial attention due to its chemical diversity and rich physical properties [19]. The crystal structure of "166" family can be viewed as the $A$-filled version of CoSn-type structure [20]. Based on the positions of $A$-site, the overall structure can be categorized into the hexagonal $HfFe_6Ge_6$-type (H-type) and hexagonal $YCo_6Ge_6$-type (Y-type, disordered structure) [21]. In the H-type structure (Fig. 1(a)), the unit cell is doubled along the $c$-axis with respect to the original CoSn-type structure and the $X$-site is also pushed away from the kagome plane by its nearby $A$-site [21]. The dimerization of the $X$-site in the kagome plane along the *c*-axis in the H-type structure is similar to the CDW-induced structural distortion in the CoSn-type FeGe [16,22]. The *A*-site can be fully occupied or of some vacancies for the H-type structure [23]. As for the case of vacancies, it forms a hybrid structure of H-type and Y-type as reported in the $R$Cr$_6$Ge$_6$ compounds [24-26]. As for the Y-type structure, the unit cell remains almost same to the CoSn-type structure, and both the *A*-site and the dimerized *X*-site have only an occupancy of 50% [27]. A quantum-limit Chern phase under magnetic fields is realized in the H-type kagome magnet TbMn$_6$Sn$_6$ [28] while CDW order transition and anomalous Hall effect have been observed in the disordered Y-type YbCo$_6$Ge$_6$ [29] and YFe$_6$Sn$_6$ [30], respectively.

Furthermore, the crystal structure of Fe-based "166" family exhibits more variants and can be either orthorhombic or hexagonal depending on the *A*-site [31]. As previously reported for the polycrystalline samples, $A$Fe$_6$Sn$_6$ with only *A* = Sc, Tm, Lu, or Zr have a hexagonal structure [23,31]. They exhibit an A-type AFM with magnetic moments along *c*-axis [23]. However, YFe$_6$Sn$_6$ single crystals are revealed to have a hexagonal crystal structure, in contrast to the reported orthorhombic structure in a polycrystalline sample [30]. Thus, the synthetic methods can be an important factor to stabilize different crystal structures of "166" family and further tailor their physical properties.

In this work, we try synthesizing single crystals of the target composition LuFe$_6$Sn$_6$ by Sn flux method. The refined structure of the crystals exhibits a large amount of vacancies on the Lu site, giving a composition of Lu$_{0.75}$Fe$_6$Sn$_6$. The structure can be viewed as a hybrid variant between HfFe$_6$Ge$_6$-type and CoSn-type. Weak ferromagnetic behavior in the *ab*-plane starts to form below 40 K under the background of A-type AFM in Lu$_{0.75}$Fe$_6$Sn$_6$. Linear magnetoresistance is also observed below 40 K for both currents applied along the *c*-axis and in the *ab*-plane in Lu$_{0.75}$Fe$_6$Sn$_6$.

## 2. Experimental details

### 2.1. Sample growth

Single crystals of Lu$_{0.75}$Fe$_6$Sn$_6$ were grown by a self-flux method. The ingredients of high-purity Lu (powder, 99.9%), Fe (powder, 99.99%), and Sn (particle, 99.9999%) with an atomic ratio of Lu: Fe: Sn = 1: 6: 20 were loaded in an alumina crucible and then sealed in an evacuated quartz tube. Subsequently, the mixture was slowly heated to 1100 °C, held at this temperature for 24 h, and then cooled down to 600 °C at a rate of 3 °C/h. Finally, the quartz tube was centrifuged at 600 °C to separate the excess Sn from the as-grown crystals. Crystals with dimensions of 1 × 1 × 3 mm$^3$ were harvested, see the inset of Fig. 1(b).

### 2.2. Structural and compositional characterization

The X-ray diffractions (XRD) on powdered and *c*-axis-aligned single crystals were performed in the range of 10° ≤ 2$\theta$ ≤ 90° by the diffractometer Bruker D2 PHASER with Cu-*K*a radiation ($\lambda$ = 1.54178 Å). The *Rietveld* refined results are further analyzed by using the General Structural Analysis System (*GSAS*) [32]. The Laue diffraction pattern with the incident X-ray parallel to the *c*-axis was recorded by back-reflection X-ray camera (Try-SE. Co., Ltd.). Single crystal X-ray diffraction (SXRD) data were collected using a Bruker D8 VENTURE with Mo-*K*α radiation ($\lambda$ = 0.71073 Å) at 293 K. The crystal structure was solved using a Superflip method [33] and refined against structure factor $F$ in Jana2006 [34]. Crystal structures were drawn by VESTA [35]. Chemical composition was ascertained through a desktop scanning electron microscope & energy-dispersive X-ray spectroscopy (EDS) (Hitachi, FlexSEM1000II).

### 2.3. Physical property measurements

Temperature-dependent magnetic susceptibility $\chi(T)$ and isothermal field-dependent magnetization $M(H)$ of single crystal Lu$_{0.75}$Fe$_6$Sn$_6$ were measured on a magnetic property measurement system with a superconducting quantum interference device (MPMS3, Quantum Design). The electrical transport (four-lead method) and heat capacity (thermal relaxation method) measurements were performed using a physical property measurement system with a model of PPMS-14 T and PPMS-9T (Quantum Design), respectively.

### 3. Results and discussions

### 3.1. Crystal structure

The target phase LuFe$_6$Sn$_6$ crystallizes in the HfFe$_6$Ge$_6$-type structure with a hexagonal space group *P*6/*mmm*, see Fig. 1(a) [23,36]. It consists of two-unit-cell

CoSn-type FeSn with Lu atoms located in the voids of only half of the Sn honeycomb planes. The other Sn atoms originally located in the kagome plane of the CoSn-type FeSn are shifted along the *c*-axis away from the inserted Lu atoms [36]. The Fe-Fe bond distance in the kagome plane is around 2.67 Å, much larger than the metallic bonding (2.45 Å) in the element Fe [37]. The direction of the *c*-axis (seen in Fig. 1(b)) can be easily identified by the shape of the hexagonal prism for the grown crystals, which is confirmed by Laue X-ray and XRD pattern on the oriented single crystals (Fig. 1(c-d)). EDS results give an average ratio of Lu: Fe: Sn = 6.09%: 46.94%: 46.97% ≈ 0.77: 6: 6, suggesting the significant amount of vacancies on the Lu-site. The Rietveld refinement of the powder XRD data (Fig. 1(e)) shows that the diffraction pattern can be well fitted by adopting the structure model of $Lu_{0.75}Fe_6Sn_6$ below with good figure of merits ($\chi^2$ = 1.417, $R_{wp}$ = 4.46%). In addition, the refined results also confirmed that there are disordered Sn-sites on the kagome plane.

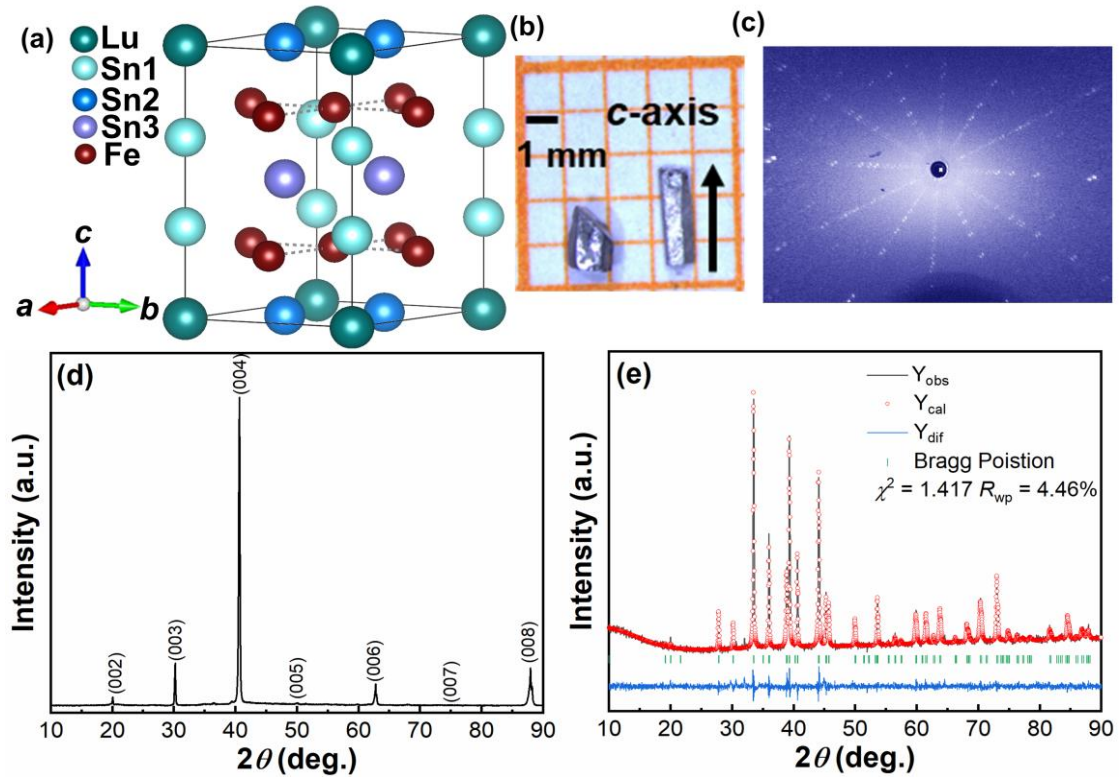

Fig. 1. (a) Crystal structure of the target phase $LuFe_6Sn_6$. (b) The photo of the grown crystals. (c-d) The Laue diffraction and XRD patterns of *c*-axis-oriented grown crystals. (e) The *Rietveld* refinement of the powder XRD data using the crystal structure of $Lu_{0.75}Fe_6Sn_6$.

To further confirm the existence of vacancies on the Lu-sites and disordered Sn-sites on the kagome plane, structure refinements were performed on the grown crystals using single crystal X-ray diffraction. The parameters for data collection and the structure refinement results are given in Table 1 and 2. The lattice parameters for the crystals are $a$ = 5.3437(1) Å and $c$ = 8.8794(2) Å, indicating a HfFe$_6$Ge$_6$-type structure [23,36]. The final $R_{obs}$ and $wR_{obs}$ values under the condition of $I > 3\sigma(I)$ are 0.0168 and 0.0376, proving a reliable fit for the collected data with around 1 e·Å$^{-3}$ residual electron or hole density. The refined occupancy of the Lu-site is 0.75 with a vacancy of 0.25, slightly less than the EDS results mentioned above. The vacancy of the Lu-site in the grown crystal is a little bit larger than the reported case in the polycrystalline samples [23], which is probably due to the different synthetic approaches. In addition, disorder at Sn1-site determined by the differential Fourier method should be included in the model which significantly reduces the $R_{obs}$ from 0.0991 to 0.0168, see Fig. 2(a). The refined occupancies of Sn1_1 and Sn1_2 are 0.81 and 0.19, respectively. Since Sn1_2 is close to the Fe kagome plane, it comes from the CoSn-type structure without the filling of Lu atoms. The disorder model at Sn1-site is also consistent with the vacancy on the Lu-site in Lu$_{0.75}$Fe$_6$Sn$_6$. Thus, the crystal structure of Lu$_{0.75}$Fe$_6$Sn$_6$ can be roughly viewed as a hybrid one between HfFe$_6$Ge$_6$-type and CoSn-type structures, see Fig. 2(b) and (c). If it is a strict model with the contributions from HfFe$_6$Ge$_6$-type and CoSn-type structures only, the occupancies of Lu and Sn1_1 should be same. We tried applying such a constrain in our model as in Ln$_x$Co$_3$(Ge$_{1-y}$Sn$_y$)$_3$ (Ln = Y, Gd) [38] and the $R_{obs}$ value increases significantly from 0.0168 to 0.025. As a result, there is a slight difference (0.75 for Lu and 0.80 for Sn1_1) in our final refined model. This small discrepancy might be due to the contribution from the partial dimerization of Sn atoms in the kagome plane along $c$ axis as in FeGe [16] and needs to be investigated in the future.

Table 1. Crystal data and structure refinement for Lu$_{0.75}$Fe$_6$Sn$_6$ single crystals at 293 K.

| | |
|---|---|
| Empirical formula | Lu$_{0.75}$Fe$_6$Sn$_6$ |
| Formula weight | 1178.7 g/mol |
| Temperature | 293 K |
| Wavelength | 0.71073 Å |
| Crystal system | hexagonal |
| Space group | P6/mmm |
| Unit cell dimensions | a = 5.3437(1) Å |
| | c = 8.8794(2) Å |
| Volume | 219.58(16) Å |
| Z | 1 |
| Density (calculated) | 8.914 g/cm$^3$ |
| Absorption coefficient | 34.577 mm$^{-1}$ |
| F(000) | 509 |
| Crystal size | 0.06 × 0.045 × 0.04 mm$^3$ |
| θ range for data collection | 2.29 to 28.29° |
| Index ranges | −7 ≤ h ≤ 7, −7 ≤ k ≤ 7, −11 ≤ l ≤ 11 |
| Reflections collected | 10915 |
| Independent reflections | 149 [$R_{int}$ = 0.0693] |
| Completeness to θ = 28.29° | 100% |
| Refinement method | F |
| Data / restraints / parameters | 149 / 0 / 18 |
| Goodness-of-fit | 2.75 |
| Final R* indices [I>3σ(I)] | $R_{obs}$ = 0.0168, $wR_{obs}$ = 0.0376 |
| R indices [all data] | $R_{all}$ = 0.0190, $wR_{all}$ = 0.0379 |
| Extinction coefficient | 0.0450(40) |
| Largest diff. peak and hole | 1.27 and −1.29 e·Å$^{-3}$ |

*R = Σ||$F_o$|−|$F_c$|| / Σ|$F_o$|, wR = {Σ[w(|$F_o$|$^2$ − |$F_c$|$^2$)$^2$] / Σ[w(|$F_o$|$^4$)$^{1/2}$ and w=1/(σ$^2$(F)+0.0001F$^2$).

Table 2. Atomic coordinates and equivalent isotropic displacement parameters (Å$^2$×10$^3$) for Lu$_{0.75}$Fe$_6$Sn$_6$ single crystals at 293 K with estimated standard deviations in parentheses.

| Label | x | y | z | Occupancy | $U_{eq}$* |
|---|---|---|---|---|---|
| Lu | 0 | 0 | 0 | 0.751(5) | 13(1) |
| Sn(3) | 1/3 | 2/3 | 1/2 | 1 | 8(1) |
| Sn(2) | 1/3 | 2/3 | 0 | 1 | 14(1) |
| Sn(1_1) | 0 | 0 | 0.3364(2) | 0.808(4) | 9(1) |
| Sn(1_2) | 0 | 0 | 0.2518(8) | 0.192 | 9(1) |
| Fe | 1/2 | 0 | 0.2465(2) | 1 | 7(1) |

*$U_{eq}$ is defined as one third of the trace of the orthogonalized $U_{ij}$ tensor.
The following occupancy constraint is added: o[Sn(1_2)] = 1 − o[Sn(1_1)].

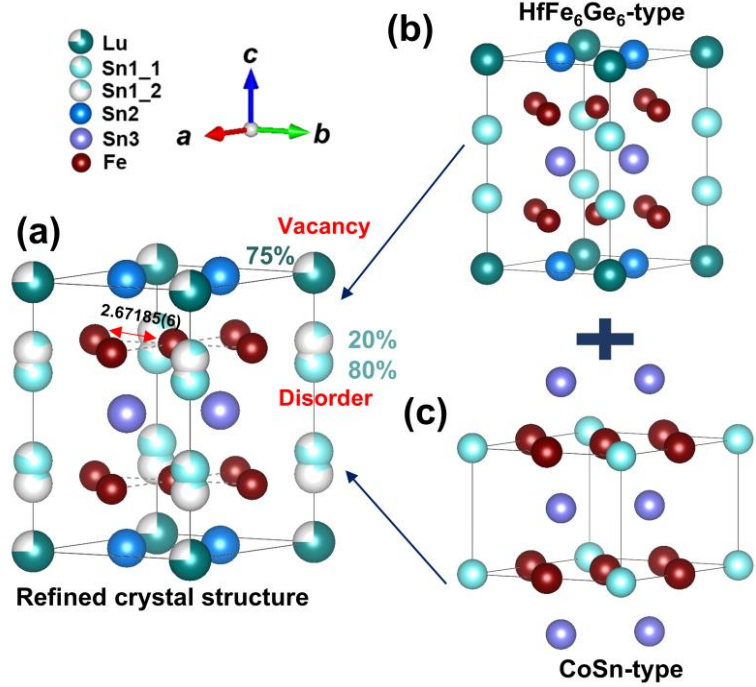

Fig. 2. (a) Refined crystal structure of the grown phase $Lu_{0.75}Fe_6Sn_6$. (b) $HfFe_6Ge_6$-type structure. (c) CoSn-type structure.

## 3.2. Magnetic property

Anisotropic temperature-dependent magnetic susceptibilities of $Lu_{0.75}Fe_6Sn_6$ under $\mu_0 H = 0.1$ T from 5 to 380 K are shown in Fig. 3(a). The magnetic susceptibility for $H \perp c$ increases rapidly below 40 K while it only shows a small upturn for $H \parallel c$, indicating a ferromagnetic-like transition with a net magnetic component in the $ab$-plane. The magnetic susceptibility for $H \parallel c$ increases with temperature from 160 up to 380 K, see the inset of Fig. 3(a), suggesting an antiferromagnetic order with a transition temperature larger than 380 K and magnetic moment along $c$-axis in $Lu_{0.75}Fe_6Sn_6$. Since $Lu_{1-\delta}Fe_6Sn_6$ polycrystalline sample exhibits an A-type AFM with magnetic moments along $c$-axis and a Néel temperature $T_N \approx 550$ K [39], $Lu_{0.75}Fe_6Sn_6$ single crystal is expected to have the same A-type antiferromagnetic order with a similar Néel temperature in spite of a small discrepancy in their stoichiometries. The small upturn in the magnetic susceptibility for $H \parallel c$ at low temperature is probably attributed to a small misalignment of the crystal during the measurement which results in a magnetic contribution from the $ab$-plane during the measurement. The magnetic susceptibility above 100 K for $H \perp c$ is well described by a modified Curie-Weiss law: $\chi = \chi_0 + C/(T-\theta)$, see Fig. 3(b), where $\chi_0$ is the temperature-independent term, $C$ is the Curie constant determined by the effective moment of the magnetic atoms, and $\theta$ is the Curie-Weiss temperature indicating the interactions between magnetic atoms. The fitted parameters are $\chi_0 = 2.72(2)\ 10^{-2}$ emu/mol, Weiss temperature $\theta = 41.05(5)$ K, and the effective moment $\mu_{eff} = 2.89(2)\ \mu_B$/Fe estimated from the following formula: $\mu_{eff} = \sqrt{\frac{3k_B C}{N_A}} \mu_B$, $k_B$ is the Boltzmann constant and $N_A$ is Avogadro constant. The positive Curie-Weiss temperature $\theta$ suggests a dominant ferromagnetic interaction among Fe atoms in $Lu_{0.75}Fe_6Sn_6$, consistent with a ferromagnetic transition below 40 K. The observed Curie-Weiss-like behavior under the background of an antiferromagnetic order might indicate an unfrozen in-plane component of magnetic moment like the similar experimental results in $Ba_{1-x}K_xMn_2As_2$ with a high doping level [30].

In order to further corroborate the ferromagnetic transition in the *ab*-plane, isothermal field-dependent magnetizations $M(H)$ at different temperatures for $Lu_{0.75}Fe_6Sn_6$ were performed with both $H \parallel$ and $\perp c$, see Fig. 3(c-d). At 150 K, magnetization is almost linear to magnetic field for both directions, in accordance with an antiferromagnetic order in $Lu_{0.75}Fe_6Sn_6$. Below 100 K, the magnetization becomes nonlinear to magnetic field and gradually shows a ferromagnetic-like behavior with decreasing the temperature for $H \perp c$. The magnetization at 5 K exhibits a typical ferromagnetic behavior with a saturated magnetic moment of around $0.2 \mu_B$ /Fe after a linear term has been subtracted. The saturated magnetic moment is also much smaller than the value $\mu_{Fe} = 2.32(4)\ \mu_B$ at 2 K determined by neutron diffraction in polycrystalline samples [23]. The small saturated magnetic moment suggests a possible canting towards the *ab*-plane and the formation of a weak ferromagnetism as observed in the $YCo_6Ge_6$-type compound $YFe_6Sn_6$ [30]. The estimated canting angle into *ab*-plane is around 5° based on the measured saturation magnetic moment by field-dependent magnetization and the magnetic moment determined by neutron diffraction. Such a scenario is also proved by Mössbauer spectroscopy which reveals the deviation of magnetic moment along *c*-axis with lowering the temperature [23]. Below 100 K, the magnetization with $H \parallel c$ starts to show a spin-flop behavior when the field is approaching to 7 T and the critical field becomes smaller with decreasing temperature, see Fig. 3(d).

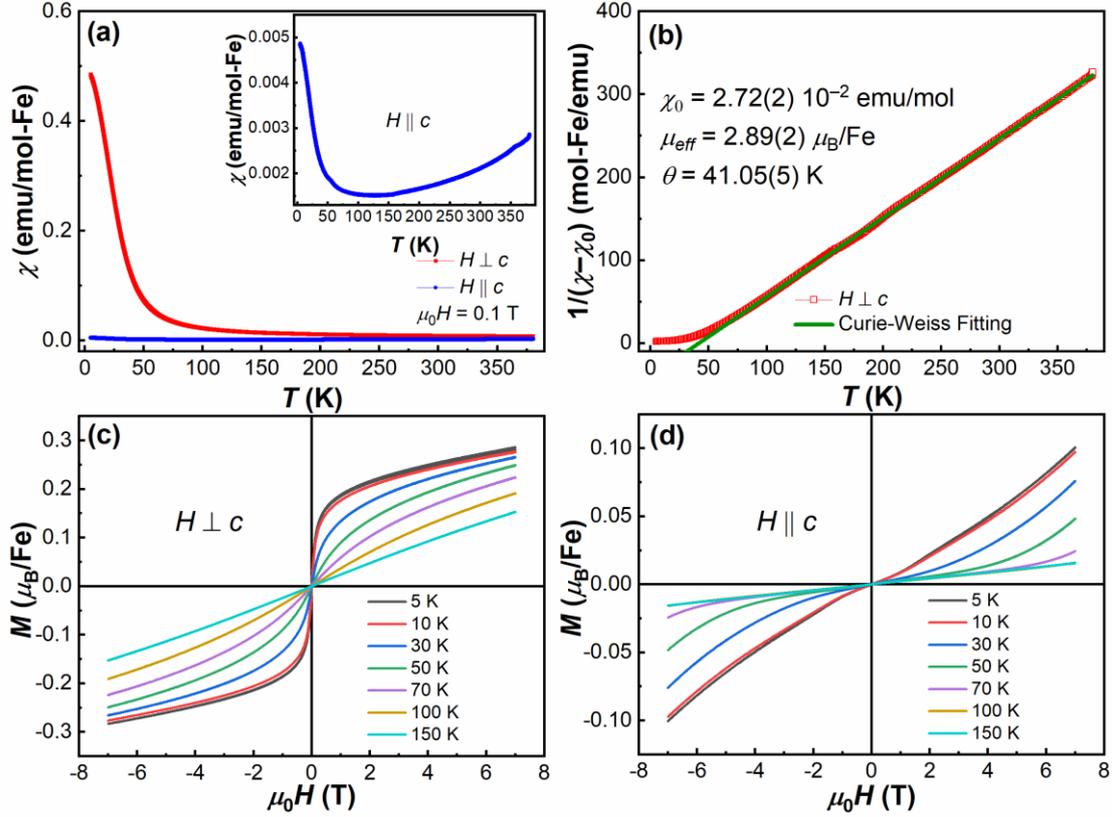

Fig. 3. (a) Temperature-dependent magnetic susceptibilities $\chi(T)$ of $Lu_{0.75}Fe_6Sn_6$ for $H \perp$ and $\parallel c$. Inset: a close-up plot for $H \parallel c$ alone. (b) Temperature-dependent inverse magnetic susceptibility $1/(\chi(T)- \chi(0))$. The solid line is the Curie-Weiss fitting with the obtained parameters. Isothermal field-dependent magnetizations $M(H)$ of $Lu_{0.75}Fe_6Sn_6$ with (c) $H \perp c$ and (d) $H \parallel c$ at different temperatures.

### 3.3. Specific heat.

Temperature-dependent specific heat of $Lu_{0.75}Fe_6Sn_6$ does not show any observable signal of a phase transition between 2 and 300 K (Fig. 4(a)), further confirming that the weak ferromagnetism probably comes from a canted AFM. The specific heat below 10 K can be well fitted (Fig. 4(b)) by the following equation:

$$C = \gamma_e T + \beta T^3 + \delta T^5$$

Here $\gamma_e T$ is the electron contribution, $\beta T^3 + \delta T^5$ is the phonon contribution [40]. The fitted values are $\gamma_e = 87.6(7)$ mJ K$^{-2}$ mol$^{-1}$, $\beta = 1.18(3)$ mJ K$^{-4}$ mol$^{-1}$ and $\delta = 5.8(3)$ $10^{-3}$ mJ K$^{-6}$ mol$^{-1}$. The Debye temperature $\Theta$ is estimated to be 273.6 K by employing the formula $(\frac{12NR\pi^4}{5\beta})^{1/3}$. The specific heat data above 250 K exceed slightly the Dulong-Petit limit $3NR$ ($N$ is the number of the atoms in the formula, and $R$ is the gas constant) [41].

Since the magnetic ordering temperature is around 550 K at $Lu_{0.75}Fe_6Sn_6$, there will be a contribution of a magnon, a collective excitation of a magnetic order at around 300 K for the specific heat data [42]. The contribution of conducting electrons at 300 K is also non-negligible. By adding these two extra terms, it is reasonable for the values of specific heat to be larger than the limit $3NR$ in $Lu_{0.75}Fe_6Sn_6$.

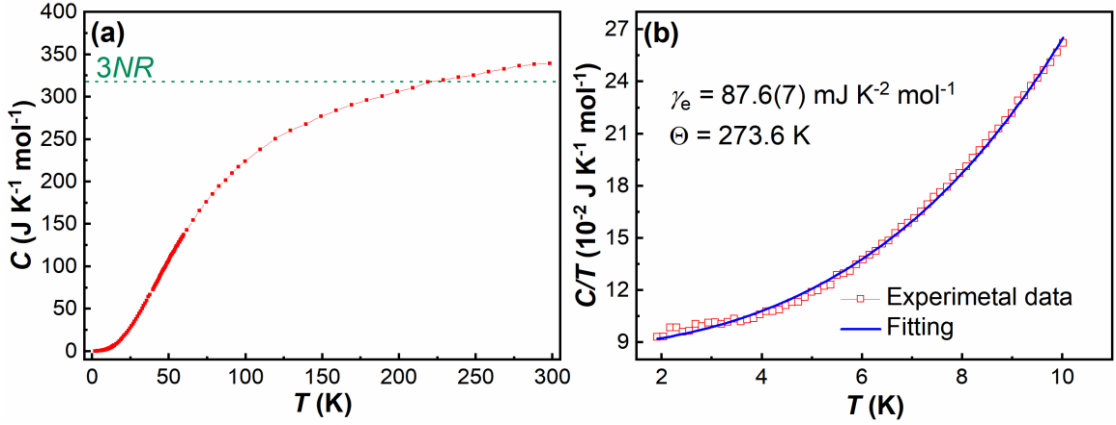

Fig. 4. (a) Temperature-dependent specific heat of $Lu_{0.75}Fe_6Sn_6$, the green dashed line is the Dulong–Petit limit. (b) The fitting of phonon and electron contributions for specific heat at low temperature.

### 3.4. Electric transport

Temperature-dependent resistivities with $I \perp$ and $\parallel c$ for $Lu_{0.75}Fe_6Sn_6$ from 2 to 300 K do not show any obvious anomaly (Fig. 5(a)). The residual resistivity ratio values ($RRR = \rho_{300\,K}/\rho_{2\,K}$) are estimated to be 2.97 and 5.28 with $I \perp$ and $\parallel c$, respectively. The low $RRR$ values originate from the disorder from the Lu and Sn1-sites inside the crystal. The derivatives of resistivity with respect to temperature from 2 to 200 K exhibit a broad hump for both directions (Fig. 5(b)), further in line with a weak ferromagnetic transition at 40 K.

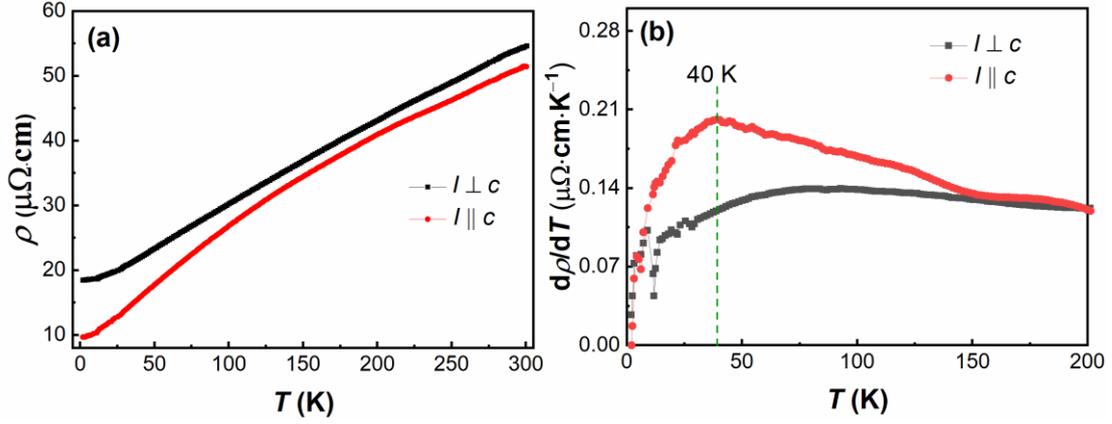

Fig. 5. (a) Temperature-dependent electric resistivities with $I \perp$ and $\parallel c$ for $Lu_{0.75}Fe_6Sn_6$. (b) The corresponding derivatives of resistivity with respect to temperature. The dashed green line is a guide for eyes.

Fig. 6(a-b) presents the anisotropic magnetoresistance ($MR = (\rho(H) - \rho(0))/\rho(0)$) of $Lu_{0.75}Fe_6Sn_6$ with the $H \parallel$ and $\perp c$. It shows a positive magnetoresistance for both directions at low temperature. The magnetoresistance is tiny at 150 K and gradually increases with decreasing temperature. Linear $MR$ with no sign of saturation is observed for both directions in $Lu_{0.75}Fe_6Sn_6$, in stark contrast to the negative $MR$ observed in $YFe_6Sn_6$ [30]. The origin of linear $MR$ can have many possible explanations such as nontrivial linear dispersion in band structure, electrons occupying the lowest spin-split Laudau level, strong sample inhomogeneity causing mobility modulations and partially gapped Fermi surfaces from density waves [43,44]. Since a significant amount of disorder in the crystal structure is revealed by the structure refinements above, the linear $MR$ in $Lu_{0.75}Fe_6Sn_6$ is probably attributed to microscopic inhomogeneities. The disorder creates the modulation of the carrier mobility and modify the current path in the sample, leading to a linear $MR$ behavior [45-47]. More experimental and theoretical investigations are needed to elucidate the origin. The $MR$ along the $c$-axis under $H \perp c$ at 5 K reaches as high as 160% under the magnetic field of 12 T while it is only 25% for the current applied in the $ab$ plane under $H \parallel c$, indicating an anisotropic effect of the disorder in regulating the $MR$ behavior. The temperature-dependent $MR$ under $\mu_0 H$ = 12 T also exhibits a rapid increase below 40 K for both directions, see Fig. 6(c-d), consistent with the weak ferromagnetic transition observed in magnetic susceptibility. Since negative $MR$ is usually present in a ferromagnetic system which reduces the magnetic scattering of the carrier under magnetic fields [48,49], the absence of such a behavior in Lu$_{0.75}$Fe$_6$Sn$_6$ suggests that disorder plays a dominant role in determining the transport behavior.

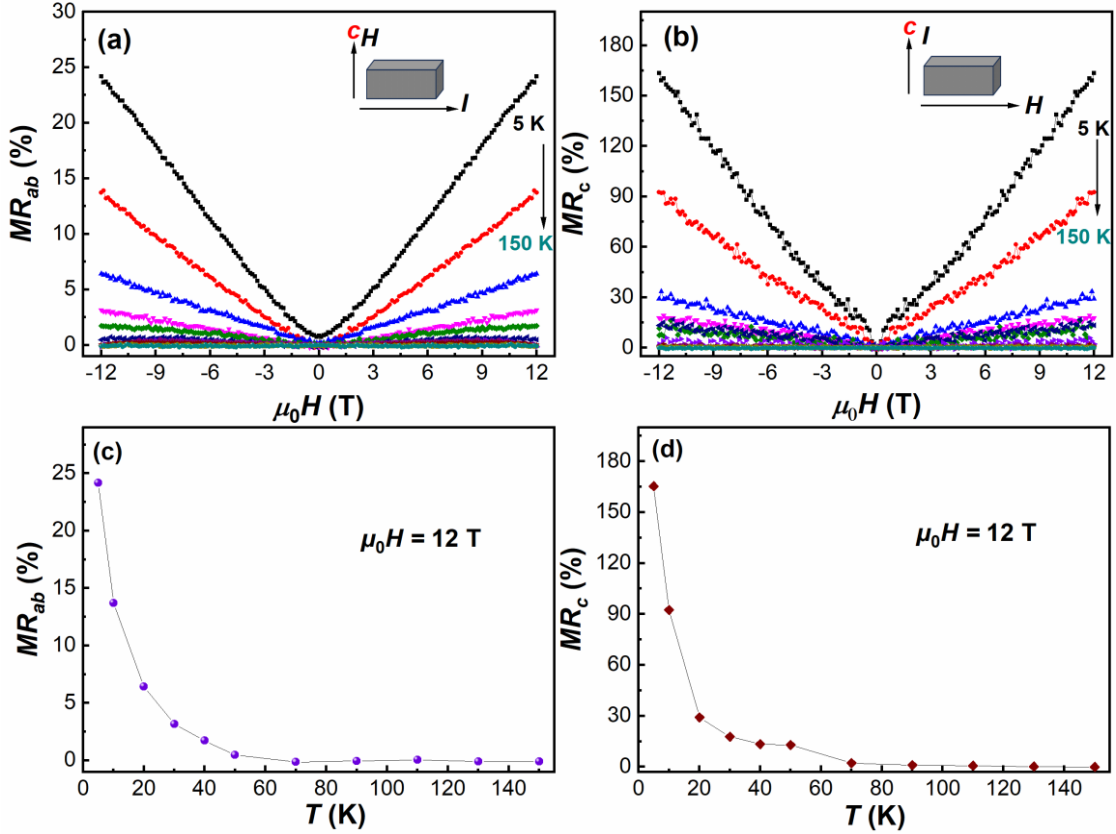

Fig. 6. Magnetoresistance at different temperatures with (a) $H \parallel c$ axis and (b) $H \perp c$ axis for Lu$_{0.75}$Fe$_6$Sn$_6$. Temperature-dependent magnetoresistance with (c) $H \parallel c$ axis and (d) $H \perp c$ axis for Lu$_{0.75}$Fe$_6$Sn$_6$ at $\mu_0 H = 12$ T.

## 4. Conclusion

In summary, single crystals of Lu$_{0.75}$Fe$_6$Sn$_6$ were synthesized by a self-flux method and its structure is resolved by single crystal X-ray diffraction. It adopts a hybrid structure of HfFe$_6$Ge$_6$-type and CoSn-type with a 25% vacancy of the Lu-site and disorder of the Sn-site in the kagome plane. Weak ferromagnetism emerges in the *ab*-plane below 40 K originating from to a canted A-type antiferromagnetism. Specific heat and electronic transport also support such a scenario. Non-saturated linear *MR* as large as 160% with the current along the *c*-axis and magnetic field in the *ab*-plane is observed in Lu$_{0.75}$Fe$_6$Sn$_6$. The *MR* behavior is probably attributed to the disorder which modulates the current flow path in the crystal. Our work provides a useful insight to study magnetism, disorder and electronic transport among the rich "166" family.

**CRediT author contributions statement**

Chenfei Shi: Investigation, Methodology, Formal analysis, Writing-original draft.

Zhaodi Lin: Methodology, Formal analysis.

Qiyuan Liu: Methodology, Formal analysis.

Junai Lv: Methodology, Formal analysis.

Xiaofan Xu: Methodology, Formal analysis.

Baojuan Kang: Methodology, Formal analysis.

Jin-hu Yang: Methodology, Formal analysis.

Yi Liu: Methodology, Formal analysis.

Jian Zhang: Methodology, Formal analysis.

Shixun Cao: Conceptualization, Validation, Supervision, Resources, Project administration, Funding acquisition.

Jin-Ke Bao: Conceptualization, Validation, Formal analysis, Supervision, Writing-review and editing, Resources, Project administration, Funding acquisition.

**Declaration of competing interest**

The authors declare no competing financial interest.

**Acknowledgements**

J.-K. B. acknowledges support from the National Natural Science Foundation of China (Grant No. 12204298), Beijing National Laboratory for Condensed Matter Physics (Grant No. 2023BNLCMPKF019), and the startup funding of Hangzhou Normal University. S. X. C. would like to acknowledge the research grant from the National Natural Science Foundation of China (Grant No. 12374116).